# Quantum entanglement in the Synchronization of Homoclinic Chaotic Spike Sequences

F.T.Arecchi [1,2*]

1- Department of Physics, Università di Firenze, Italy.
2 - INO-CNR,Largo E.Fermi, 6 -50125- Firenze, Italy.
*e-mail: tito.arecchi@ino.it

**Abstract**

Physics deals with Newtonian particles described by position $q$ and momentum $p$. The precision of the simultaneous measurement of $q$ and $p$ is limited by the uncertainty relation ruled by Planck's constant. From the uncertainty relation all quantum consequences emerge, including ***entanglement***.

On the other hand, Homoclinic Chaos (*HC*) , that consists of sequences of identical pulses, unevenly spaced in time, entails a non-Newtonian description. Synchronization of finite *HC* spike sequences (*SFSS*) display quantum features ruled by a constant different from ℏ, yielding entanglement ..

As a relevant example, we describe how brain neurons generate *HC* voltage pulses.; *SFSS* is the way two different words coded as *HC* pulses compare their content and extract a meaningful sequence by exploiting quantum entanglement that lasts over a de-coherence time in the range of human linguistic processes.

**MAIN**

Newtonian dynamics describes a particle via a coordinate $q$ and momentum $p$. Both $q$ and $p$ assume continuous values; however, their joint measurement is restricted by the quantum uncertainty relation. This is the basis of all quantum phenomena, including entanglement.

A case of non-Newtonian dynamics is Shilnikov Homoclinic Chaos (*HC)* [1,2], observed in chemical [3] and laser [4] systems. *HC* consists of a sequence of identical spikes separated by temporal bins *τ*, each bin labeled as ***1*** (*spike present*) or ***0*** (*spike absent*). (Fig.1).

In the case of brain neurons (Fig.2), the minimal spike separation is around 3 *ms*. In human language operations, comparison of a word with a previous one entails the synchronization of finite neuronal spike sequences (*SFSS*) [5,6].

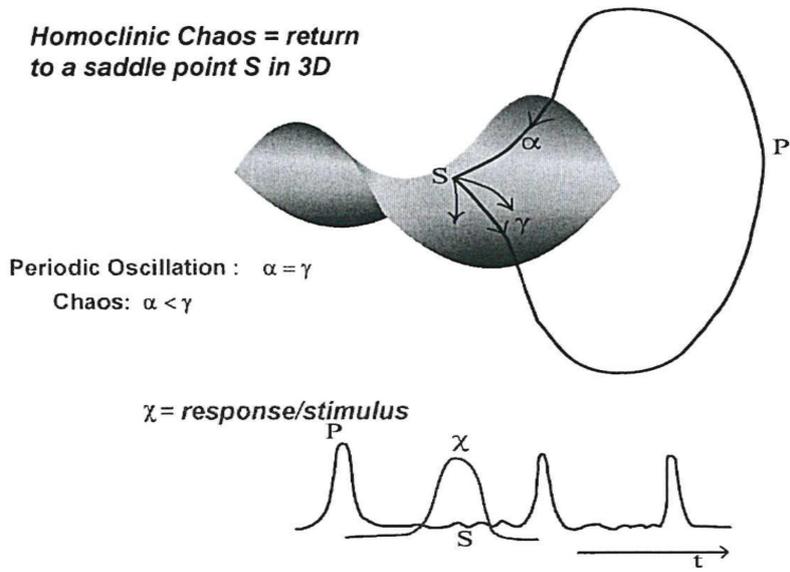

Fig.1-*HC yields closed trajectories that return to a saddle point S with an approach rate $\alpha$ smaller than the escape rate $\gamma$ ($\alpha < \gamma$). Representing the position along the trajectory as a vertical signal versus time, the paths yield identical pulses P (spikes) separated by a time interval depending on the permanence time around the saddle. We call such an orbit **homoclinic** to stress its return to the same saddle. The permanence time around S can be controlled by an external stimulus, thus, the homoclinic sequence can be synchronized to an applied signal.*

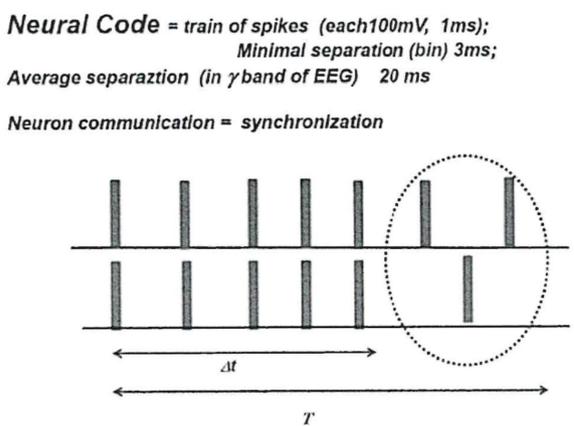

*Fig. 2-Brain neuron activity consists of standard electric spikes 100 mV high, with a minimal mutual separation of 3 ms (HC dynamics). Example of coupling of two neurons by synchronized spike trains; synchronization missed after $\Delta t$ for an extra-spike in the upper train. In the case of two spike trains of duration T, synchronized up to $\Delta T$, the number of different realizations is $2^{(T-\Delta T)}$.*

Information –time uncertainty (*ITU*) expressed as spike number - time uncertainty, represents a quantum limitation for *HC*. Because of *ITU*, the comparison of two sequences of spikes by *SFSS* entails entangled spike sequences *(ES)*,. Thus, entanglement explains how different words connect in a linguistic task. An example is reported in Fig.3.

*ITU* provides an explanation for other quantum effects in human cognitive processes lacking a plausible framework, since either no appropriate quantum constant had been associated [7], or speculating on processes ruled by Planck's constant resulted in unrealistic de-coherence times.[8].

*ES* provides a computational speed up in the comparison of a word with a previous one (meaning problem), thus connecting two pieces of a linguistic sequence. The meaning problem consists of finding the best congruence between a word just occurred in the last reading of a linguistic text and an associated word occurring in the previous piece recalled via the short term memory. The process is confined within a de-coherence time, that is the time allotted for meaning extraction.

We outline the dynamical processes in single neurons and how neurons couple via *SFSS*.

i) Within a neuron, voltage travels via an axon as a soliton spike with the energy loss replaced by extra energy provided by $10^7$ ionic channels, each one entailing an ATP →ADP+P chemical reaction that releases 0.3 *eV* [9].

ii) In single neurons, *HC* yields identical 100 *mV* spikes of 1 *ms* duration, with minimal separation *(bin)* $\tau$= 3 *ms*, non-uniformly distributed in time [10]; the inter-play between gamma and theta bands of EEG contributes to tailoring the sequence length [11, 12].

iii) Neurons get into speaking terms by synchronizing their spike sequences over a finite time interval (*SFSS= synchronization of finite spike sequences*) [13, 14].

iv) The role of *SFSS* is exploited by the *GWS = global workspace*. In fact, *GWS* was introduced [15, 16] to yield the most appropriate motor reaction. Here, we extend the *GWS* role by postulating that, acting as a synchronization reader, it extracts the word matching.

Let us explore in detail the above processes.

*HC* consists of a homoclinic orbit in a 3D space[10], returning to a saddle S with an approach rate $\alpha$ and an escape rate $\gamma$ (Fig.1). If $\gamma > \alpha$, then the orbital period is chaotic, otherwise it is regular. The 3D orbit yields a standard spike voltage (neural case: 100 *mV*, 1 *ms*, repeating in time with a minimal separation *(bin)* $\tau = 3$ *ms*) ( Fig.2 ). The height of the saddle S , that drives the escape time and hence the inter-spike separation, is affected by the voltage provided by the neighbouring cortical areas via EEG signals. Spikes are clustered around each EEG peak and then absent up to the next peak; this yields clusters of spikes separated by wide empty inter-spike intervals. The main EEG signal is in the gamma band (40 to 80 *Hz*), The gamma band is further modulated by a lower frequency theta band (around 7 *Hz*) that pushes down some of the gamma maxima, thus introducing new gaps in the spike sequence [11, 12].

Synchronization of the spike sequences over a finite time (*SFSS*) results from the interplay between sensorial stimuli and the local potential provided by the combination of gamma and theta bands. In case of competition between two different cortical areas providing different amounts of synchronization, the Global Work Space (*GWS*) reads the corresponding spike sequences and decides the most appropriate one to trigger motor actions [15, 16] . In the linguistic operation, we have postulated that *GWS* output provides word matching.

Two separate moments characterize human cognition, namely, *apprehension (A)* (duration around 1 *s* [17] ) whereby a coherent perception emerges from neuronal groups stimulated by sensorial stimuli, yielding a motor reaction, and *judgment (B),* whereby-under a *(A)* stimulus- memory recalls a previous *(A)* unit coded in a suitable language and the comparison of the two *A* units yields a meaning ,.

The linguistic operations entails the comparison between two *(A)*'s acquired at different times, the previous one recalled by the memory. *SFSS* extracts the conformity of the second one, on the basis of the first one. If a word has *N* different meanings, the most appropriate one is that with the largest synchronization with the next piece. In linguistic performances, by the theta-gamma modulation, the spike train coding the first word is interrupted from a duration *T* to *ΔT<T*. To get synchronization, it must be lengthened by *T−ΔT* ; this can occur in *N=2(T−ΔT)* ways by filling the *T−ΔT* interval with *N* different sequences of *0* and *1*. Thus the first word is coded as a cluster state *|E>* consisting of *N* different sequences.

Let e.g. *ΔT=T−1*, then synchronization of *|T>* with *|ΔT>* amounts to comparing *|T>* with the entangled state

$$|E> = 1/\sqrt{2} \; ( \; |\Delta T, 0 > + |\Delta T, 1>) \qquad (1)$$

Without further treatment, the most synchronized state will be identical to the second world. If however the states are each weighted differently by a semantic operator *€* (*€* stays for **emotion**) [18,19], then the emerging most synchronized word is the joint result of *€* and of the code of the second world (Fig.3).

We call *T* the time the brain takes to build up a decision. *T* corresponds to the reaction time for visual lexical decisions or word naming. It occurs in a range from 300 to 700 *ms* [5, 17]. Then, the total number of binary words that can be processed is $P_M = 2^{T/\tau}$. If e.g. *T =300 ms*, it follows $P_M \approx 10^{33}$.

Interruption of a spike train introduces a joint uncertainty in the word information and spike duration (*ITU*). Let us investigate what mechanisms rule the duration time. The threshold for spike onset is modulated by the *EEG gamma* oscillation. Phase coherence of the gamma wave over distant regions permits spike synchronization overcoming delays due to the finite propagation speed in the axons . Furthermore, the lower frequency, *EEG theta band*, controls the number of gamma maxima within a processing interval. The theta-gamma cross-modulation corresponds to stopping the neural sequence at *ΔT≤T* [20, 21]. As a result, all spike trains equal up to *ΔT*, but different by at least one spike in the interval *T−ΔT*, provide an uncertainty cloud
$\Delta P = 2^{(T-\Delta T)/\tau} = P_M 2^{-\Delta T/\tau}$ , where $P_M = 2^{T/\tau}$.

Thus, an uncertainty of exponential type holds between spike information *P* and duration *T*,

$$\Delta P \cdot 2^{\Delta T/\tau} = P_M \qquad (2),$$

The whole *T* train is a vector of a $2^T$ dimensional space; synchronization of two different *T* trains amounts to counting the number of *0* and *1* coincidences. Since synchronization entails equal lengths of the trains under comparison, a *ΔT* pulse acquires a length *T* by being entangled with all possible sequences of *0's* and *1's* in the complementary interval *T−ΔT*. As we project state *|T>* over *| E>,* we extract the congruence as the best synchronization S, namely
  S= < E| T>   (0<S>1) .

As an example, take *T=10, ΔT=9*; hence, the following synchronization values result:

$S_1=(9+1)/10=1; \; S_2=(9+0)/10=0.9.$

Define a fractional bit number $u=P/P_M$; then the fractional uncertainty $\Delta u=\Delta P/P_M$ is related to the gated time $\Delta T$ by

$$\Delta u \cdot 2^{\Delta T/\tau}=1 \qquad (3)$$

The two conjugated quantities, $u$ and $\Delta T$, are coupled by an exponential type uncertainty. By a change of variable $y=\tau 2^{t/\tau}$, we arrive to a product type uncertainty relation

$$\Delta u \Delta y = \tau \qquad (4)$$

Thus, in the space $(u,y)$ we have a Heisenberg-like uncertainty relation.

We stress the difference between Newtonian physics and *HC*. In Newtonian physics, $q$ and $p$ can vary over unlimited intervals, provided their joint measurement be constrained by an uncertainty given by Planck's constant. On the contrary, *HC* has a lower bound corresponding to one spike present/absent within a time bin. In neuronal *HC*, the associated uncertainty constant $C$ is expressed in *Joules ×sec*, once we explicit the energy per spike. This corresponds to the opening along the axon of $10^7$ ionic channels [9] each one entailing an *ATP →ADP+P* chemical reaction yielding 0.3 *eV*, thus the minimal energy disturbance in neural spike dynamics is $3 \cdot 10^{-13}$ *J*, that is, around $10^8 k_B T_r$ ($k_B$ Boltzmann constant, *Tr* room temperature ). Since $\tau=3ms$, it results $C \approx 10^{-15} Js \approx 10^{19} \hbar$.

However, due to the structure of Eq. (2), the uncertainty holds over a finite range, between two extremes, namely (measuring times in $\tau$ units)

i)     $\Delta T_{min}=1$,    yielding   $\Delta P_{max}=P_M/2$

and

ii)     $\Delta T_{max}=T$,    yielding   $\Delta P_{min}=1$.

Based on standard quantum approach, we expect two-sequence entanglement within such a time interval. The entangled state lasts over a ***de-coherence time***. For $\Delta P=1$ (minimal disturbance), the de-coherence time is

$$\Delta y_d = P_M \tau \qquad (5)$$

Using the numbers reported above, $\tau_d$=*de-coherence time*= 0.3*s*, very far from $\tau_d=\hbar/k_B T_r \approx 10^{-14} s$ evaluated for single Newtonian particles disturbed by the thermal energy $k_B T_r$ [8].

Notice that the resulting $\tau_d$ is equal to the full processing time $T=300$ *ms* chosen as an example. If we consider a different processing time for the *SFSS*, the de-coherence time changes accordingly.

Fig.3 visualizes the linguistic operations in a simple case

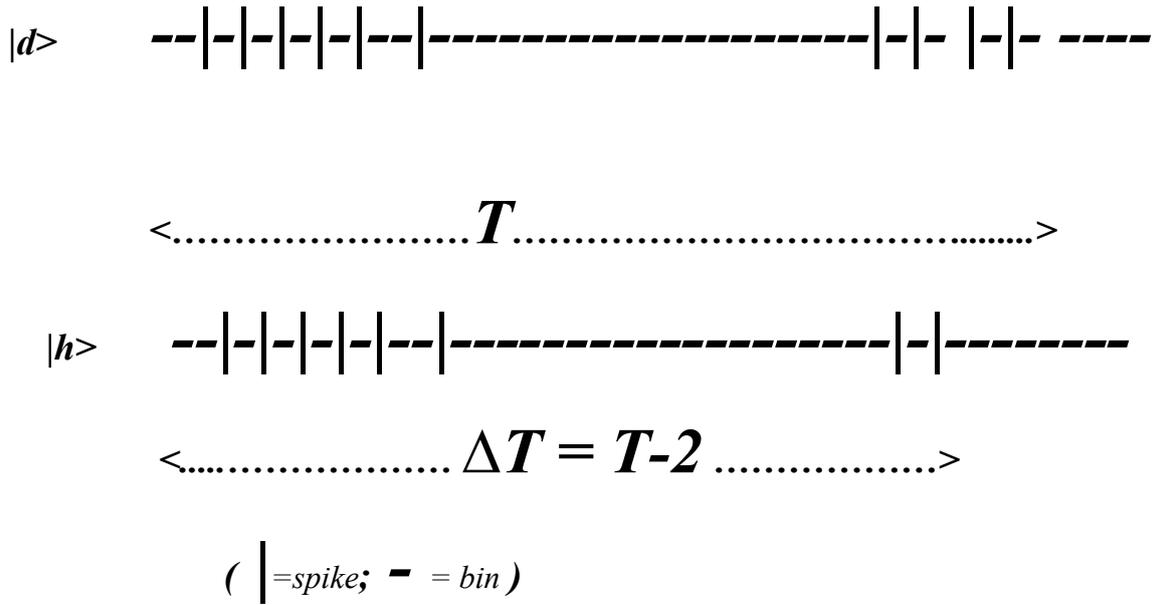

*( | =spike; — = bin )*

*Fig3 - Example of a linguistic process. The word of the second piece is |d> = |T>, and the word of the first piece |h> = |ΔT> is interrupted at ΔT = T-2 .*

*Report the first word to T bits: |h> = |ΔT> + ¼ (|00> + |01> + |10> + |11>.*

*Now, if the last two bits of |d> are |10> , then synchronization of |h> with |d> leads to*

| States in T-ΔT | Synchro. <h\|d> |
|---|---|
| 00 | 0 |
| 01 | 0 |
| 10 | 1 |
| 11 | 0.5 |

→ *max. congruence for |h>=|ΔT> +|10>, that coincides with |d> : **trivial !***

*If however, before synchronization we apply to |h> the operator € with the following weights*

| States in T-ΔT | weight € |
|---|---|
| 00 | 1/4 |
| 01 | 1/4 |
| 10 | 0 |
| 11 | 1/2 |

*the combined effect of € and S recovers the state |h'> = |ΔT> + |11>. Thus,, the sequence: i) ΔT (interruption); ii) € (different weights for states in interval T-ΔT ; iii) S (synchronization), yields the novelty |h'> ≠ |h>.*

We now introduce interference in time, rather than in space as in Newtonian quantum mechanics. We introduce a wave-like assumption

$$\omega = 1/\tau. \; P/P_M = u/\tau \qquad (6)$$

If $N$ is the total spike number over time $T$ and $N'$ the spike number in the interrupted interval $\Delta T$, then

$$u = P/P_M = 2^{N'}/2^N = 2^{-(N-N')} = 2^{-(T-\Delta T)/\tau} \qquad (7)$$

The equivalent of the two slit self-interference of a single particle would be the comparison of a single spike train of $P$ bits with a delayed version of itself. An implementation would consist of the spike train translated in time and superposed to the original train. As one changes the time separation $\Delta t$ of the two trains, the spike synchronization (number of coincidences of 1's) decays from $N$ (number of 1's in the spike train) to $\sqrt{N}$ (random overlaps).
However, further increasing the time separation, self-interference entails a **revival** of the synchronization depending on the Fourier periodicity, that is, for

$$\Delta t = \tau P_M/P = \tau 2^{(T-\Delta T)/\tau} = \tau 2^x \qquad (8)$$

where $x = (T-\Delta T)/\tau$ is the normalized time lapse between the whole train and the interrupted version. Thus, to have revivals, the time translation $y = \Delta t/\tau$ must be larger than the time lapse $x$. Comparing two interrupted sequences with lapses respectively $x$ and $x'$, we generate two interferential returns corresponding respectively to $x$ and $x'$ and thus separated by $(x'-x)$. The wave character, that in particle dynamics is associated with $k = p/\hbar$, here is due to the duration of the sequence to be synchronized with the initial reference sequence of duration $T$. Thus, it is bound to the theta – gamma cross modulation.

We apply the formalism to a linguistic task, consisting of the comparison of two words, one corresponding to the last presentation, and the previous one recovered by the short memory within 2-3 $s$ [22]. Words are coded as trains of neuronal spikes. Performance of the linguistic task amounts to synchronization of the two trains. Take $T$ as the time duration of the second word. The previous one is interrupted at $\Delta T < T$ by theta-gamma modulation. Such a word spans a region of a finite-dimensional space. The total spike train belongs to a space of $2T$-dimensions and is represented by $|T\rangle$. The spike train interrupted after a duration $\Delta T$ provides a set of states living in the same $2T$-dimensional space. It is entangled with all possible realizations of $1s$ and $0s$ in the complementary interval $T-\Delta T$.
For sake of reasoning, in Eq.(1) we considered the minimal interruption, $\Delta T = T-1$. In such a case, the synchronization task of $|T\rangle$ with $|\Delta T\rangle$ amounts to comparing $|T\rangle$ with the entangled state (1) and then performing a measurement based quantum computation [23]. In general, $T-\Delta T = N$, hence synchronization amounts to measuring over a cluster of $2N$ entangled states, that is, comparing the whole train of duration $T$ with $2N$ different interrupted versions of it, each one displaying differences from the original *T* train and hence defeating full synchronization. Thus, a word recognition consists of the comparison between a reference word living in *2T* dimensions, and a tentative word retrieved via the short term memory and interrupted to *ΔT* by the theta-gamma EEG modulation. If we compare this process with measurement-based quantum computation [23], there, once a cluster (*h*) has been prepared, its component states must be coupled by some interaction. As a fact, the theta-gamma modulation creates the cluster (*h*) of entangled states; but the successive synchronization *S* selects state |*d*> and thus it would be an irrelevant operation. We postulate that – before *S* is applied - the entangled states are coupled by emotional operators $\epsilon$ [18,19]. Thus, our SFSS- operated language behavior consists of the sequence:

i) the interruption *ΔT* yields *2(T−ΔT)* entangled states |*h*>;

ii) each one of these states is modified by the emotional coupling ($\epsilon$) as
$\epsilon$|*h*> →|*h\**>;

iii) synchronization *S* selects the state |*h\**> that best synchronizes to |*d*>(congruence).

Without ii), the choice of |*h*> due to |*d*> would be a trivial operation. We have seen that quantizing the spike train implies a time interruption. As a fact, spikes occur at average rates corresponding to the EEG gamma band. However, the lower frequency theta band controls the number of gamma band bursts. For instance, gamma power in the hippocampus is modulated by the phase of theta oscillations during working memory retention, and the strength of this cross-frequency coupling predicts individual working memory performance [20, 21].

We here postulate that emotional effects raised by the first piece of a linguistic text induce a theta band interruption of the gamma band bursts, thus introducing an entanglement that speeds up the exploration of the semantic space in search of the meanings that best mutually match. In this behaviour, emotions do not have an aesthetic value "per se", as maintained by neuro-aesthetic approaches [24], but rather they provide a fast scanning of all possible meanings within a de-coherence time. Hence, the final decision does not depend on the emotions raised by the single word, but it is the result of the comparison of two successive pieces of a linguistic sequence.

We summarize the linguistic endeavour. A previous piece of a text **(h)** is retrieved by the short term memory, modified by emotions ($\epsilon$) into **(h\*)** and compared via synchronization (S) with the next piece **(d).**

We stress the revolution brought about by *HC*, in fact, *SFSS* cannot be grasped in terms of position-momentum variables; thus, the quantum constant for spike train position-duration uncertainty is not Planck's constant.

The minimal energy disturbance which rules the de-coherence time is not $k_B T_R$ ($T_R$ room temperature); rather, since it corresponds to *ΔP=1*, it entails the minimal energy necessary to add or destroy a cortical spike. This energy corresponds to the opening along the axon of about $10^7$ ionic channels each one requiring an *ATP →ADP+P* reaction involving 0.3 *eV* [9], thus the minimal energy disturbance in neural spike dynamics is around $10^8 k_B T_R$.

This is the evolutionary advantage of *HC*, that is, the brain lives at room temperature $T_R$ and yet is barely disturbed, as it were cooled at $10^{-8} T_R$.[25].

The procedure here introduced explains other reported evidences of quantum effects in human cognitive processes, so far lacking a plausible framework. Models of quantum behavior in language and decision taking have already been considered [26-28]. The speculations introduced to justify a quantum behaviour can be grouped in two categories, namely, either

    i)     they lack a dynamical basis and thus do not consider limitations due to a quantum constant [26], hence, they do not inquire for a de-coherence terminating the operation;

or

    ii)     they refer to Newtonian particles [27, 28] and hence are limited by a de-coherence time around $10^{-14} s$, well below the times of cognitive processes.

In conclusion, while in the *perceptual* case the cognitive action combines a bottom-up signal provided by the sensorial organs with a top-down interpretation provided by long term memories stored in extra-cortical areas, in the *linguistic* case the comparison occurs between the code of the second piece and the code of the previous one retrieved by the short term memory. In this case, theta –gamma modulation introduces an information-time uncertainty (*ITU*), hence an entanglement among different words that provides a fast quantum search of meanings. In *SFSS,* the associated de-coherence time is compatible with the processing times of linguistic endeavors. On the contrary, all previously reported approaches either overlook the need for a quantization constant , or they quantized Newtonian particles and consequently displayed very short de-coherence times.

**REFERENCES**

1.Shilnikov,L.P.: A case of the existence of a countable set of periodic motions ,*Sov. Math.Dokl.***6,** 163-166 (1965)

2. Shilnikov,L.P.: On a new type od bifurcation if multi-dimensional dynamical systems,

    *Sov. Math.Dokl.***10**, 1368 -1371(1970)

3.Arneodo,A., Coullet P., Spiegel E.A. , and Tresser C., Homoclinic Chaos in chemical systems, *Physica D,***14**, 327  (1985)

4. Arecchi F.T., Lapucci A., Meucci R., Roversi J.A., and Coullet P.H.: Experimental characterization of Shilnikov chaos by statistics of return times , *Europhys. Lett.***6**, 677-682(1988)

5. P.J. Uhlhaas, G. Pipa, B. Lima, L. Meloni, S. Neuschwander, D. Nikolic and W. Singer Neural synchrony in cortical networks: history, concept an current status. *Frontiers in Integrative Neuroscience,* **3,** (17) ,1-19 (2009)

6. W. Singer, Consciousness and the structure of neuronal representations. *Phil. Trans. R. Soc. Lond.* **353**, 1829 (2015)

7. J. R Busemeyer and P.D. Bruza.. *Quantum models of cognition and decision*. Cambridge University Press, (2012)

8. M. Tegmark: Importance of quantum de-coherence in brain processes,. *Phys. Rev.E,* **61**,4194–4206 (2000)

9. D. Debanne, E. Campanac, A. Bialowas, E. Carlier, and G. Alcaraz., Axon Physiology, *Physiological Reviews*, **91**(2),555–602, (2011)

10. F. T. Arecchi., Chaotic neuron dynamics, synchronization and feature binding, *Physica A* **338** , 218- 237 (2004)

11. O. Jensen and L. L. Colgin,. Cross-frequency coupling between neuronal oscillations, *Trends in Cognitive Sciences*, **11**(7),267–269, (2007)

12. J.E. Lisman and O. Jensen., The Theta-Gamma Neural Code, *Neuron*, **77**(6),1002-1016 (2013)

13. C. von der Malsburg,. *The correlation theory of brain function*. - Internal Report 81-2, Dept. of Neurobiology, Max-Planck-Inst. for Biophysical Chemistry, Göttingen, Germany (1981)

14. W. Singer and C.M. Gray. Visual Feature Integration and the Temporal Correlation Hypothesis, *Annu. Rev. Neurosci.*, **18**(1):555–586, (1995)

15. B.J. Baars *A cognitive theory of consciousness*. Cambridge University Press,. (1993)

16. S. Dehaene, C. Sergent, and J.P.Changeux. A neuronal network model linking subjective reports and objective physiological data during conscious perception, *Proc Natl. Acad. Sci. U S A*, **100**,8520 (2003)

17. E. Rodriguez, N. George, J-P. Lachaux, J. Martinerie, B. Renault, and F. J Varela,: Perception's shadow: long-distance synchronization of human brain activity, *Nature*, **397**,430– 433 ( 1999)

18. Damasio, A R. and Carvalho, G.B.. The nature of feelings: Evolutionary and neurobiological origins. *Nature Reviews. Neuroscience.* **14** (2): 143–52. (2013)

19. Damasio A. R. *The feeling of what happens: Body and emotion in the making of consciousness*. Random House (2000)

20. P.Fries,: A mechanism for cognitive dynamics: neuronal communication through neuronal coherence*, Trends in Cognitive Science* , **9** (10), 474-480 (2005)

21. P. Fries, D. Nikolic and W. Singer, The gamma cycle, *Trends in Neurosciences*, **30** (7), 309-316 (2007)

22. E. Poeppel: Lost in time: a historical frame, elementary processing units and the 3-second window, *Acta neurobiologiae experimentalis*, **64**(3):295–302 ,(2004)

23. Briegel, H. J. , Browne D.E., Dür W., Raussendorf R., and Van den Nest M. Measurement based quantum computation, *Nature Physics*, **5**(1):19–26 (2009)..

24. S. Zeki and J. Nash. *Inner vision: An exploration of art and the brain*, vol. 415. Oxford University Press , Oxford, (1999).

25. F.T.Arecchi, F.T, Dynamics of consciousness: complexity and creativity, *The Journal of Psychophysiology*,**24** (2),141-148. (2010)

26. R. W. Gibbs Jr. *Embodiment and cognitive science*. Cambridge University Press (2005).

27. R. Penrose.: *Shadows of the Mind*, OxfordUniversity Press Oxford ,(1994)

28. S. Hagan, S. R Hameroff, and J. A Tuszyński:. Quantum computation in brain microtubules: De-coherence and biological feasibility*, Phys. Rev.E*,**65**, 061901 (2002)